\documentclass[12pt]{article}
\usepackage{epsfig,amssymb,amsmath,psfrag,dsfont}

\def \bl  {\begin{align*}}
\def \el  {\end{align*}}

\def \be  {\begin{equation}}
\def \ee  {\end{equation}}
\def \ba  {\begin{eqnarray}}
\def \ea  {\end{eqnarray}}
\def \baa {\begin{eqnarray*}}
\def \eaa {\end{eqnarray*}}
\def \bb  {\begin {thebibliography} }
\def \eb  {\end{thebibliography}}
\def \lab #1 {\label{#1}}

\def \qqquad {\qquad\quad}

\def \ra  {\rightarrow}

\newcommand{\nn}{\nonumber}
\newcommand{\beq}{\begin{equation}}
\newcommand{\eeq}{\end{equation}}
\newcommand{\beqa}{\begin{eqnarray}}
\newcommand{\eeqa}{\end{eqnarray}}
\newcommand{\da}{\dot\alpha}
\newcommand{\db}{\dot\beta}

\newcommand{\tlam}{\tilde{\lambda}}

\def \e  {\mathop{\rm e}\nolimits}

\renewcommand{\a}{\alpha}
\newcommand{\adt}{{\dot{\alpha}}}
\newcommand{\bdt}{{\dot{\beta}}}
\renewcommand{\b}{\beta}

\def\e{\epsilon}
\def\d{\delta}
\def\lam{\lambda}
\def\tlam{\tilde{\lambda}}
\def\th{\theta}
\def\cN{\mathcal{N}}
\def\cP{\mathcal{P}}
\def\cW{\mathcal{W}}
\def\cZ{\mathcal{Z}}
\def\AA{\mathcal{A}}
\def\BB{\mathcal{B}}
\def\CC{\mathcal{C}}
\def\DD{\mathcal{D}}
\def\EE{\mathcal{E}}
\def\FF{\mathcal{F}}
\def\GG{\mathcal{G}}
\def\HH{\mathcal{H}}
\def\MM{\mathcal{M}}
\def\OO{\mathcal{O}}
\def\UU{\mathcal{U}}
\def\NN{\mathcal{N}}

\def\del{\partial}

\def\l<{\langle}
\def\r>{\rangle}

\def\XXint#1#2#3{{\setbox0=\hbox{$#1{#2#3}{\int}$}
     \vcenter{\hbox{$#2#3$}}\kern-.5\wd0}}

\renewcommand{\title}[1]{\vbox{\center\LARGE{#1}}\vspace{5mm}}
\renewcommand{\author}[1]{\vbox{\center#1}\vspace{5mm}}

\begin{document}

\thispagestyle{empty}
\null\vskip-12pt \hfill  LAPTH-001/10 \\
\vskip2.2truecm
\begin{center}
\vskip 0.2truecm {\Large\bf
{\Large Yangians, Grassmannians and T-duality}
}\\
\vskip 1truecm
{\bf J.~M. Drummond and L. Ferro \\
}

\vskip 0.4truecm
{\it
LAPTH\footnote{Laboratoire d'Annecy-le-Vieux de Physique Th\'{e}orique, UMR 5108}, Universit\'{e} de Savoie, CNRS\\
B.P. 110,  F-74941 Annecy-le-Vieux Cedex, France\\
\vskip .2truecm                        }
\end{center}

\vskip 1truecm 
\centerline{\bf Abstract} 
We investigate the Yangian symmetry of scattering amplitudes in $\NN=4$ super Yang-Mills theory and show that its formulations in twistor and momentum twistor space can be interchanged. In particular we show that the full symmetry can be thought of as the Yangian of the dual superconformal algebra, annihilating the amplitude with the MHV part factored out.
The equivalence of this picture with the one where the ordinary superconformal symmetry is thought of as fundamental is an algebraic expression of T-duality.
Motivated by this, we analyse some recently proposed formulas, which reproduce different contributions to amplitudes through a Grassmannian integral. We prove their Yangian invariance by directly applying the generators.

\medskip

 \noindent

\newpage
\setcounter{page}{1}\setcounter{footnote}{0}

\section{Introduction}

Maximally supersymmetric Yang-Mills theory exhibits many remarkable properties. It is a superconformal quantum field theory which is widely believed to be equivalent to a supersymmetric string theory on the background $AdS_5 \times {S}^5$ \cite{Maldacena:1997re,Gubser:1998bc,Witten:1998qj}. Furthermore, in studying its planar limit, many advances have been made which point towards the existence of an underlying integrable structure which governs the behaviour of the various physical quantities in the theory. Great progress has been made on the spectral problem of anomalous dimensions of gauge-invariant operators (see e.g. \cite{AdS/INT1,AdS/INT1b}) where various techniques from the field of integrable systems have been applied, extending previous work in QCD \cite{Lipatov:1993yb,Faddeev:1994zg}.

Scattering amplitudes in planar $\cN=4$ super Yang-Mills theory are also constrained by hidden symmetries. In particular one can consider a dual coordinate space, related to the particle momenta via $p_i = x_i - x_{i+1}$. In fact it turns out that amplitudes are related to Wilson loops on the light-like polygonal contour with cusps located at the dual points $x_i$. This occurs both in the strong coupling regime \cite{Alday:2007hr} and, for MHV amplitudes, in the perturbative regime \cite{Drummond:2007aua,Brandhuber:2007yx,Drummond:2007cf,Drummond:2007au,Drummond:2007bm,Drummond:2008aq}.
The fact that amplitudes are related to Wilson loops in the dual space implies that the conformal symmetry of the Wilson loops also acts on amplitudes. This new dual conformal symmetry is distinct from the original conformal symmetry of the Lagrangian. As was shown in \cite{Drummond:2008vq} it extends naturally to a dual superconformal symmetry which partially overlaps with the original superconformal symmetry.

On tree-level amplitudes both the original and the dual superconformal symmetries are unbroken (except on singular kinematical configurations \cite{Bargheer:2009qu, Korchemsky:2009hm,Sever:2009aa}). The breaking of the original superconformal symmetry  by loop corrections  is still not completely understood (see recent discussions in \cite{Korchemsky:2009hm,Sever:2009aa}), while the breaking of the dual conformal symmetry is under control and it is identified with the breaking of the conformal symmetry of the Wilson loop in the dual space (in recent papers \cite{Alday:2009zm, Henn:2010bk} a different regularisation has been used in which the symmetry is unbroken).

It was shown in \cite{Drummond:2009fd} that the combination of the original superconformal and dual superconformal symmetries forms a Yangian structure in the bilocal representation described in \cite{Dolan:2003uh,Dolan:2004ps}. The original superconformal symmetry can be thought of as the `level-zero' superconformal subalgebra inside the Yangian while the non-trivial dual superconformal generators provide part of the bilocal `level-one' generators. The full Yangian can then be generated by taking commutators of this set of generators. The Yangian can be thought of as the quantisation of the loop algebra of the superconformal algebra which arises as the full symmetry group of the classical $AdS$ sigma model \cite{Bena:2003wd}. This integrable structure can be thought to arise from the fact that the full supersymmetric background maps into itself under a combination of bosonic and fermionic T-dualities \cite{Berkovits:2008ic,Beisert:2008iq,Beisert:2009cs}.

Recently some remarkable formulas have been proposed which reproduce many different contributions to amplitudes. The idea is to take an integral over a Grassmannian of certain superconformally invariant delta functions \cite{ArkaniHamed:2009dn}.
In fact it was conjectured in \cite{ArkaniHamed:2009dn} that every object obtained by choosing some integration contour for the Grassmannian integral is a leading singularity of an $\cN =4$ super Yang-Mills amplitude. If this conjecture is true then one can obtain different terms in the BCFW expansion of the tree-level amplitudes, box-integral coefficients appearing in one-loop amplitudes or, more generally, higher-loop leading singularities.

A very similar formula was proposed in \cite{Mason:2009qx} but where this time the delta functions are written in terms of the momentum twistors introduced in \cite{Hodges:2009hk}. This makes the dual superconformal properties of the formula manifest and again it turns out that the integrations yield integral coefficients for amplitudes. In fact the equivalence of the two formulas was demonstrated in \cite{ArkaniHamed:2009vw}  through a change of variables, therefore showing that both symmetries are present in the Grassmannian integral. The objects it produces are thus Yangian invariants. For recent progress on identifying the various expressions produced in this way see \cite{Bullimore:2009cb,Kaplan:2009mh,ArkaniHamed:2009dg}.

In this paper we will show that the interchange between the original and momentum twistor formulations can be seen as an algebraic feature of the Yangian $Y(psu(2,2|4))$. Specifically we will show that there is an equivalent (T-dual) formulation of the Yangian symmetry where the dual superconformal symmetry plays the role of the level-zero subalgebra and the original superconformal generators provide some of the level-one generators, again in a bilocal representation. This fact is the algebraic expression of the T self-duality of the $AdS$ sigma model discussed in \cite{Berkovits:2008ic,Beisert:2008iq,Beisert:2009cs}.
We will then show that the Yangian generators can be used to provide a very direct proof of the invariance of the Grassmannian formulas.

The paper is organised as follows. In section \ref{sect-amplitudes} we describe the on-shell superspace description of scattering amplitudes in $\mathcal{N}=4$ super Yang-Mills theory. In section \ref{sect-symmetries} we describe the superconformal and dual superconformal symmetries exhibited by tree-level amplitudes and recall the fact that these symmetries form a Yangian symmetry. In section \ref{sect-T-dual} we describe the alternative (T-dual) representation of the symmetry. Then in section \ref{sect-grassmannians} we recall the basic structure of the Grassmannian formulas of \cite{ArkaniHamed:2009dn,Mason:2009qx} and finally in section \ref{sect-inv} we show how the Yangian generators can be used to show the Yangian invariance of the Grassmannian formulas directly.

\section{On-shell scattering amplitudes}
\label{sect-amplitudes}

The on-shell supermultiplet of $\cN=4$ super Yang-Mills theory is conveniently described by a superfield $\Phi$, dependent on Grassmann parameters $\eta^A$ which transform in the fundamental representation of $su(4)$. The on-shell superfield can be expanded as follows
\be
\Phi = G^+ + \eta^A \Gamma_A + \tfrac{1}{2!} \eta^A \eta^B S_{AB} + \tfrac{1}{3!} \eta^A \eta^B \eta^C \e_{ABCD} \overline{\Gamma}^D + \tfrac{1}{4!} \eta^A \eta^B \eta^C \eta^D \e_{ABCD} G^-.
\label{onshellmultiplet}
\ee
Here $G^+,\Gamma_A,S_{AB}=\tfrac{1}{2}\e_{ABCD}\overline{S}^{CD},\overline{\Gamma}^A,G^-$ are the positive helicity gluon, gluino, scalar, anti-gluino and negative helicity gluon states respectively. Each state $\phi \in \{G^+,\Gamma_A,S_{AB},\overline{\Gamma}^A,G^-\}$ carries a definite on-shell momentum
\be
p^{\a \adt} = \lam^\a \tlam^\adt,
\ee
and a definite weight $h$ (called helicity) under the rescaling
\be
\lam \longrightarrow \a \lam, \qquad \tlam \longrightarrow \a^{-1} \tlam, \qquad \phi(\lam,\tlam) \longrightarrow \a^{-2h} \phi(\lam,\tlam).
\ee
The helicities of the states appearing in (\ref{onshellmultiplet}) are $\{+1,+\tfrac{1}{2},0,-\tfrac{1}{2},-1\}$ respectively. If, in addition, we assign $\eta$ to transform in the same way as $\tlam$,
\be
\eta^A \longrightarrow \a^{-1} \eta^A,
\ee
then the whole superfield $\Phi$ has helicity 1. In other words the helicity generator,
\be
h = -\tfrac{1}{2} \lam^\a \frac{\del}{\del \lam^\a} + \tfrac{1}{2} \tlam^\adt \frac{\del}{\del \tlam^\adt} + \tfrac{1}{2} \eta^A \frac{\del}{\del \eta^A},
\ee
acts on $\Phi$ in the following way,
\be
h \Phi = \Phi.
\ee
When we consider scattering amplitudes\footnote{We refer throughout the paper to colour-ordered amplitudes.} of the on-shell superfields then we have that the helicity condition (or `homogeneity condition') is satisfied for each particle, i.e.
\be
h_i \mathcal{A}(\Phi_1,\ldots,\Phi_n) = \mathcal{A}(\Phi_1,\ldots,\Phi_n), \qquad i=1,\ldots,n.
\label{Ahelicity}
\ee
The tree-level amplitudes in $\cN=4$ super Yang-Mills theory can be written as follows,
\be
\mathcal{A}(\Phi_1,\ldots,\Phi_n)=\mathcal{A}_n =  \frac{\d^4(p) \d^8(q)}{\l<12\r> \ldots \l<n1\r>} \cP_n(\lam_i,\tlam_i,\eta_i) = \mathcal{A}_n^{\rm MHV} \mathcal{P}_n.
\label{amp}
\ee
The MHV tree-level amplitude,
\be
\mathcal{A}_n^{\rm MHV} = \frac{\d^4(p) \d^8(q)}{\l<12\r> \ldots \l<n1\r>},
\ee
contains the delta functions $\d^4(p) \d^8(q)$ which are a consequence of translation invariance and supersymmetry and it can be factored out leaving behind a function with no helicity,
\be
h_i \mathcal{P}_n = 0, \qquad i=1,\ldots,n.
\label{Phelicity}
\ee
The explicit form of the function $\mathcal{P}_n$ which encodes all tree-level amplitudes was found in \cite{Drummond:2008cr} by solving a supersymmetrised version \cite{Brandhuber:2008pf,ArkaniHamed:2008gz,Elvang:2008na} of the BCFW recursion relations \cite{Britto:2004ap,Britto:2005fq}.

Beyond tree-level, the function $\mathcal{P}_n$ is infrared divergent and so, as well as the kinematical dependence, necessarily has some dependence on the infrared regularisation. The general structure of the function is a sum of transcendental integral functions $F_I$ (which contain infrared divergences) multiplied by rational coefficients $c_I$, where $I$ labels the different integral topologies,
\be
\mathcal{P}_n  = \sum_{I} c_I F_I.
\label{integralexpansion}
\ee
At one loop a basis for the relevant integral functions comes from the scalar box integrals \cite{Bern:1994zx}. The tree-level amplitude is necessarily a particular linear combination of the one-loop box function coefficients due to consistency with the condition of infrared factorisation \cite{Bern:2004bt}. Other coefficients at one-loop, the four-mass box coefficients, do not appear at tree-level as the corresponding integrals are infrared finite. The one-loop coefficients $c_I$ can be determined by comparing the discontinuities of the amplitude with those of the scalar box integrals \cite{Bern:1994zx,Bern:1994cg,Britto:2004nc}.
Beyond one loop there are many more integral topologies which can contribute to the amplitude. Nonetheless the coefficients can be determined again by comparing the discontinuities of the amplitude and the integrals.

\section{Symmetries}
\label{sect-symmetries}

Maximally supersymmetric Yang-Mills is a superconformal field theory so we should expect that this is reflected in the structure of the scattering amplitudes. Indeed the space of functions of the variables $\{\lam_i,\tlam_i,\eta_i\}$ admits a representation of the superconformal algebra \cite{Witten:2003nn}, given in the appendix (\ref{shortderiv}). From the algebraic relations (\ref{comm-rel}) one finds that the algebra is generically $su(2,2|4)$ with central charge $c = \sum_i (1-h_i)$. Amplitudes are in the space of functions with helicity 1 for each particle so we have that $c=0$ after imposing the helicity conditions (\ref{Ahelicity}) and the algebra acting on the space of homogeneous functions becomes $psu(2,2|4)$.

At tree-level there are no infrared divergences and amplitudes are annihilated by the generators of the standard superconformal symmetry (up to contact terms which vanish for generic configurations of the external momenta, see \cite{Bargheer:2009qu,Korchemsky:2009hm,Sever:2009aa}),
\be
j_a \mathcal{A}_n = 0. \label{scs}
\ee
Here we use the notation $j_a$ for any generator of the superconformal algebra $psu(2,2|4)$,
\be
j_a \in \{p^{\a\adt},q^{\a A}, \bar{q}^{\adt}_A,m_{\a\b}, \bar{m}_{\adt\bdt},r^A{}_B,d,s^\a_A,\bar{s}_{\adt}^A,k_{\a \adt} \}.
\ee
The explicit form of the generators acting on the on-shell superspace coordinates $(\lam_i,\tlam_i,\eta_i)$ is given in the appendix. In fact the superconformal symmetry holds term by term in the BCFW expansion of the tree-level amplitudes. The invariance was shown directly by applying the generators to the explicit form of the amplitudes in \cite{Witten:2003nn} for MHV amplitudes and \cite{Korchemsky:2009jv} for NMHV amplitudes.

In addition the amplitudes also obey dual superconformal symmetry \cite{Drummond:2008vq}. This is best revealed by defining dual variables,
\be
x_i^{\a \adt} - x_{i+1}^{\a \adt} = \lam_i^\a \tlam_i^\adt, \qqquad \th_i^{\a A} - \th_{i+1}^{\a A} = \lam_i^\a \eta_i^A.
\label{dualvars}
\ee
Dual superconformal symmetry acts canonically on the dual superspace variables $x_i,\th_i$. It also acts on the on-shell superspace variables in order to be compatible with the defining relations (\ref{dualvars}). The form of the dual superconformal generators is given in (\ref{dualsc}).

The amplitudes can be expressed in the dual variables by eliminating ($\tlam_i,\eta_i$) in favour of $(x_i,\th_i)$. Then we have
\be
\mathcal{A}_n = \frac{\d^4(x_1-x_{n+1}) \d^8(\th_1-\th_{n+1})}{\l<12\r>\ldots \l<n1\r>} \cP_n(x_i,\th_i),
\ee
and the amplitudes are covariant under certain generators of the dual superconformal algebra defined in \cite{Drummond:2008vq}. Explicitly, it was conjectured in \cite{Drummond:2008vq} that
\begin{align}
K^{\a \adt} \mathcal{A}_n &= -\sum_i x_i^{\a\adt} \mathcal{A}_n \notag\\
S^{\a A} \mathcal{A}_n &= -\sum_i \theta_i^{\a A} \mathcal{A}_n \notag \\
D \mathcal{A}_n &= n \mathcal{A}_n,
\label{dualcov}
\end{align}
with remaining generators of the dual superconformal algebra annihilating the amplitudes. This conjecture was shown to hold  in \cite{Brandhuber:2008pf}, using the supersymmetric BCFW recursion relations.
In addition the dual superconformal algebra has a central charge $C=\sum_i h_i$ which is equal to $n$ on the space of homogeneous functions\footnote{i.e. functions satisfying the homogeneity condition (\ref{Ahelicity}).}.

In order to put the dual superconformal symmetry on the same footing as invariance under the standard superconformal algebra (\ref{scs}), the covariance (\ref{dualcov}) can be rephrased as an invariance of $\mathcal{A}_n$ by a simple redefinition of the generators \cite{Drummond:2009fd},
\begin{align}
K'^{\a \adt} &= K^{\a \adt} + \sum_i x_i^{\a \adt}, \\
S'^{\a A} &= S^{\a A} + \sum_i \theta_i^{\a A}, \\
D' &= D - n.
\end{align}
The redefined generators still satisfy the commutation relations of the superconformal algebra, but now with vanishing central charge, $C' = 0$.
Then dual superconformal symmetry is simply
\be
J'_a \mathcal{A}_n = 0.
\ee
Here we use the notation $J'_a$ for any generator of the {\sl dual} copy of $psu(2,2|4)$,
\be
J'_a \in \{P_{\a\adt},Q_{\a A}, \bar{Q}_{\adt}^A,M_{\a\b}, \overline{M}_{\adt\bdt},R^A{}_B,D',S_\a^{\prime A},\overline{S}^{\adt}_A,K'^{\a \adt} \}.
\ee

In order to have both symmetries acting on the same space it is useful to restrict the dual superconformal generators to act only on the on-shell superspace variables $(\lam_i,\tlam_i,\eta_i)$. Then one finds that the generators $P_{\a \adt},Q_{\a A}$ become trivial while the generators $\{\bar{Q},M,\bar{M},R,D',\bar{S}\}$ coincide (up to signs) with generators of the standard superconformal symmetry. The non-trivial generators which are not part of the $j_a$ are $K'$ and $S'$. In \cite{Drummond:2009fd} it was shown that the generators $j_a$ and $S'$ (or $K'$) together generate the Yangian of the superconformal algebra, $Y(psu(2,2|4))$. The generators $j_a$ form the level-zero $psu(2,2|4)$ subalgebra\footnote{We use the symbol $[O_1,O_2]$ to denote the bracket of the Lie superalgebra, $[O_2,O_1] = (-1)^{1+|O_1||O_2|}[O_1,O_2]$.},
\be
[j_a,j_b] = f_{ab}{}^{c} j_c.
\ee
In addition there are level-one generators $j_a^{(1)}$ which transform in the adjoint under the level-zero generators,
\be
[j_a,j_b\!{}^{(1)}] = f_{ab}{}^{c} j_c\!{}^{(1)}.
\ee
Higher commutators among the generators are constrained by the Serre relation\footnote{The symbol $\{\cdot,\cdot,\cdot\}$ denotes the graded symmetriser.},
\begin{align}
&[j^{(1)}_a , [j^{(1)}_b,j_c]] + (-1)^{|a|(|b| + |c|)} [j^{(1)}_b,[j^{(1)}_c,j_a]] + (-1)^{|c|(|a|+|b|)} [j^{(1)}_c,[j^{(1)}_a,j_b]] \notag \\
&= h^2 (-1)^{|r||m|+|t||n|}\{j_l,j_m,j_n\} f_{ar}{}^{l} f_{bs}{}^{m} f_{ct}{}^{n} f^{rst}.
\end{align}
The level-zero generators are represented by a sum over single particle generators,
\be
j_a = \sum_{k=1}^n j_{ka}.
\label{levelzero}
\ee
The level-one generators are represented by the bilocal formula,
\be
j_a\!{}^{(1)} = f_{a}{}^{cb} \sum_{k<k'} j_{kb} j_{k'c}.
\label{bilocal}
\ee
Thus finally the full symmetry of the tree-level amplitudes can be rephrased as
\be
y \mathcal{A}_n = 0,
\ee
for any $y \in Y(psu(2,2|4))$.

\section{T-dual representation of the symmetries}
\label{sect-T-dual}

In this section we want to show that there is an alternative (T-dual) representation of the symmetry where it is the dual superconformal generators which play the role of the level-zero generators and the additional non-trivial generators of the standard superconformal symmetry which generate the rest.
We recall that in the representation of the Yangian (\ref{levelzero},\ref{bilocal}) there was no room for the generators of dual translations $P_{\a \adt}$ and dual supertranslations $Q_{\a A}$. These generators were trivialised by restricting to the on-shell superspace (where they do not act at all). The analogous step in the dual representation of the Yangian will be to trivialise the corresponding generators of the standard superconformal algebra $p^{\a\adt},q^{\a A}$. We will achieve this by working on the support of the delta functions in (\ref{amp}) where these generators become zero. In fact we will factor out the full MHV tree-level amplitude so that we are looking at functions with zero helicity in all particles. We are thus looking at symmetries of the function $\mathcal{P}_n$ rather than the amplitude $\mathcal{A}_n$.
Then dual superconformal symmetry becomes
\be
J_a \mathcal{P}_n = 0.
\ee
To work out the consequences of the ordinary superconformal symmetry for the function $\mathcal{P}_n$ we need to use the following \cite{Witten:2003nn},
\begin{align}
0=k_{\alpha \dot\alpha} \mathcal{A}_n=k_{\a \adt} \frac{\d^4(p) \d^8(q)}{\l<12\r>\ldots\l<n1\r>} \mathcal{P}_n(\lam_i,\tlam_i,\eta_i)&= \sum_{i=1}^n \frac{\del^2}{\del \lam_i^\a \del \tlam_i^{\adt}}\frac{\d^4(p) \d^8(q)}{\l<12\r>\ldots\l<n1\r>} \mathcal{P}_n(\lam_i,\tlam_i,\eta_i)\\
&= \d^4(p) \d^8(q) \sum_{i=1}^n \frac{\del^2}{\del \lam_i^\a \del \tlam_i^{\adt}} \frac{\mathcal{P}_n(\lam_i,\tlam_i,\eta_i)}{\l<12\r>\ldots\l<n1\r>} \label{secondstep} \\
&= \d^4(p) \d^8(q) \sum_{i=1}^{n-1} \frac{\del^2}{\del \lam_i^\a \del \tlam_i^{\adt}} \frac{\mathcal{P}_n(\lam_i,\tlam_i,\eta_i)}{\l<12\r>\ldots\l<n1\r>}. \label{thirdstep}
\end{align}
To obtain the second equality (\ref{secondstep}) one needs to use the fact that we have
\be
J_a \mathcal{P}_n = 0,
\ee
in particular for the generators $J_a \in \{M_{\a \b}, \overline{M}_{\adt \bdt}, D, \bar{Q}_{\adt}^A\}$. The third equality (\ref{thirdstep}) follows from the fact that (super) amplitudes have a definite helicity ($h_i=1$) for each external particle and hence we can write the function $\mathcal{P}_n$ so that it does not depend on $p_n$ (or similarly $q_n$),
\be
\mathcal{A}_n = \frac{\delta^4(p) \delta^8(q)}{\l<12\r>\ldots\l<n1\r>} \mathcal{P}_n(\lam_i,\tlam_i,\eta_i) = \frac{\delta^4(p) \delta^8(q)}{\l<12\r> \ldots \l<n1\r>} \mathcal{P}_n(p_1,\ldots,p_{n-1},q_1,\ldots,q_{n-1}).
\ee
From (\ref{thirdstep}) we deduce\footnote{Here and throughout the paper we assume generic values for the kinematical variables and so are ignoring any contact terms which appear in the action of $\frac{\partial}{\partial \tilde\lambda}$ on $\frac{1}{\langle i i+1 \rangle}$.}
\beq
\sum_{i=1}^{n-1} \Bigl[\frac{\del}{\del \lam_i^\a} \frac{1}{\l<12\r>\ldots\l<n1\r>} \frac{\del}{\del \tlam_i^{\adt}} + \frac{1}{\l<12\r>\ldots\l<n1\r>} \frac{\del^2}{\del \lambda_i^\a \del \tilde\lambda_i^\adt} \Bigr]\mathcal{P}_n(\lam_i,\tlam_i,\eta_i) = 0,
\eeq
and hence we have that
\be
k'_{\a \adt} \mathcal{P}_n = 0,
\label{Pinv}
\ee
where
\be
k'_{\a \adt} = \sum_{i=1}^{n-1} \Bigl[\Bigl(\frac{\lambda_{i-1 \,\, \a}}{\langle i-1\, i\rangle} - \frac{\lambda_{i+1 \,\, \a}}{\langle i \, i+1 \rangle}\Bigr)\frac{\del}{\del \tilde\lambda_i^{\adt}} + \frac{\del^2}{\del \lambda_i^\a \del \tilde\lambda_i^\adt} \Bigr] .
\label{k'}
\ee

Thus we find a second order operator $k'$ which annihilates $\mathcal{P}_n$. We could now express this in terms of the variables $x_i$ and $\theta_i$ however it turns out that it is very convenient to make a further change of variables and express this operator, as well as the dual superconformal generators $J_a$, in terms of momentum (super)twistors. These variables parametrise the twistor space associated with the dual space with coordinates ($x_i,\theta_i$). They were recently introduced in \cite{Hodges:2009hk} to give a geometrical interpretation of the cancellation of spurious singularities in tree-level amplitudes.

Momentum twistors $\cW_i^{\AA} = (\lam_i^\a,\mu_i^\adt,\chi_i^A)$ are defined in terms of the dual variables $x_i$ and $\theta_i$ by the following relations,
\be
\mu_i^{\adt} = x_i^{\a \adt} \lam_{i \a}, \qqquad \chi_i^A = \th_i^{\a A}\lam_{i \a}.
\label{momtw}
\ee
When expressed in terms of the momentum twistors the dual superconformal generators $J_a$ are almost identical in form to the original superconformal generators $j_a$ expressed in terms of the ordinary twistors. For example we have
\begin{align}
&P_{\a \adt} = \sum_i \lambda_{i \a} \frac{\del}{\del \mu_i^{\adt}}, &&Q_{\a A} = \sum_i \lam_{i\a} \frac{\del}{\del \chi_i^A} \notag\\
&\overline{Q}_{\adt}^A = \sum_i \chi_i^A \frac{\del}{\del \mu_i^{\adt}}, && D = -\sum_i \Bigl[\frac{3}{2} \mu_i^{\da}\frac{\partial}{\partial\mu_i^{\da}} + \chi_i^A \frac{\partial}{\partial\chi_i^{A}} + \frac{1}{2} \lambda_{i}^{\alpha}\frac{\partial}{\partial\lambda_i^{\alpha}} \Bigr] \notag\\
& M_{\alpha\beta} = \sum_i \lambda_{i(\alpha} \frac{\partial}{\partial\lambda_i^{\beta)}}, &&
\overline{M}_{\da\db} = \sum_i \mu_{i(\da} \frac{\partial}{\partial\mu_i^{\db)}}.
\label{dualgen}
\end{align}
The full set of generators can be written in terms of the momentum supertwistors as\footnote{When we write e.g. $(-1)^{\AA +\CC}$ then $\AA$ and $\CC$ are shorthand for the gradings of the indices $\AA$ and $\CC$, namely $0$ for a bosonic index and $1$ for a fermionic one, and addition is always understood to be mod $2$.}
\be
J^{\AA}{}_{\BB} = \sum_i \Bigl[\cW_i^{\AA} \frac{\del}{\del \cW_i^{\BB}} - \tfrac{1}{8} (-1)^{\AA+\CC} \delta^{\AA}_{\BB} \cW_i^{\CC} \frac{\del}{\del \cW_i^{\CC}}\Bigr].
\ee
We will usually write this formula without the second term, with the removal of the supertrace to be understood. Also we note that the helicity conditions (\ref{Phelicity}) become
\be
\left[ \lambda_{i}^{\alpha}\frac{\partial}{\partial\lambda_i^{\alpha}} + \mu_i^{\da}\frac{\partial}{\partial\mu_i^{\da}} + \chi_i^A\frac{\partial}{\partial\chi_i^{A}} \right] \mathcal{P}_n= \cW_i^{\AA} \frac{\partial}{\partial \cW_i^{\AA}} \mathcal{P}_n = 0
\ee
in terms of the momentum twistor variables.

We would now like to show that invariance given by the operator $k'$ is equivalent to level-one generators given by the same  bilocal formula (\ref{bilocal}) but now in terms of the dual superconformal densities $J_{ia}$. In other words we would like to show that the operators
\be
J^{(1)}_a = f_{a}{}^{cb} \sum_{i<j} J_{ib} J_{jc}
\label{dualbilocal}
\ee
annihilate $\mathcal{P}_n$. To do so we will follow a similar analysis to that in \cite{Drummond:2009fd} and identify $k'$ with $P^{(1)}$ up to terms which themselves annihilate $\mathcal{P}_n$.

For the generator $P^{(1)}$, the bilocal formula (\ref{dualbilocal}) in the dual representation of the Yangian symmetry reads
\beq
P_{\alpha\da}^{(1)} = \sum_{i<j} \left[ M_{i\alpha}^{\gamma} P_{j\gamma\da} + \overline{M}_{i\da}^{~\db} P_{j\alpha\db} - D_i P_{j\alpha\da} + \overline{Q}_{\da i}^{C} Q_{j\alpha C} - (i\leftrightarrow j) \right].
\label{Pone}
\eeq
To show the equivalence of this generator to $k'$ when acting on $\mathcal{P}_n$, we take the expression $(\ref{Pinv})$
and use the chain rule to pass to the momentum supertwistor variables,
\beqa
\frac{\partial}{\partial \lambda_i^{\alpha}} &\longrightarrow&  \frac{\partial}{\partial \lambda_i^{\alpha}} + \sum_k \frac{\partial \mu_k^{\adt}}{\partial \lambda_i^{\alpha}} \frac{\partial}{\partial\mu_k^{\adt}} + \sum_k \frac{\partial \chi_k^{A}}{\partial \lambda_i^{\alpha}} \frac{\partial}{\partial\chi_k^{A}} , \\
\frac{\partial}{\partial \tlam_i^{\da}} &\longrightarrow&  \sum_k \frac{\partial \mu_k^{\bdt}}{\partial \tlam_i^{\da}} \frac{\partial}{\partial\mu_k^{\db}}.
\label{chain}
\eeqa
To see that these are the correct relations one must remember that the momentum twistor variables (\ref{momtw}) depend on the on-shell variables $\lambda_i,\tilde\lambda_i$ both explicitly and implicitly through the dual superspace coordinates $x_i,\theta_i$. Specifically we have
\beqa
\mu_k^{\da} &=& x_1^{\alpha\da} \lambda_{k\alpha} - \sum_{j=1}^{k-1} \langle jk \rangle ~\tlam_j^{\da}, \nn \\
\chi_k^{A} &=& \theta_1^{\alpha A} \lambda_{k\alpha} - \sum_{j=1}^{k-1} \langle jk \rangle ~\eta_j^{A}.
\eeqa
The coefficients of the $\mu$ and $\chi$ derivatives in (\ref{chain}) then follow from these relations.

Performing the change of variables in (\ref{chain}) we find that the first order term in (\ref{k'}) becomes
\be
-\sum_{i<j} \Bigl( \frac{\lambda_{i-1 \alpha}}{\langle i-1 ~i\rangle} - \frac{\lambda_{i+1 \alpha}}{\langle i ~i+1\rangle} \Bigr) \lambda_i^{\gamma} \lambda_{j \gamma} \frac{\del}{\del \mu_j^{\adt}}
\ee
which can be rewritten as
\be
-\sum_{i<j} \Bigl( \frac{\lambda_{i-1 \alpha} \lambda_i^{\gamma}}{\langle i-1 ~i\rangle} - \frac{\lambda_{i+1}^{\gamma}\lambda_{i\alpha}}{\langle i ~i+1\rangle} - \delta_\alpha^\gamma\Bigr)  \lambda_{j \gamma} \frac{\del}{\del \mu_j^{\adt}} ~.
\label{firstorderterm}
\ee
Since the first two terms under the sum differ by one step in $i$, they cancel pairwise leaving the first with $i=1$ and the second with $i=j-1$. The latter term is zero, being proportional to $\langle j j \rangle$ while the former can be written as
\be
-\sum_{j=1}^n \frac{\lambda_{n\a} \lambda_1^\gamma}{\langle n1 \rangle} \lambda_{j\gamma} \frac{\del}{\del \mu_j^\adt} = - \frac{\lambda_{n\a} \lambda_1^\gamma}{\langle n1 \rangle} P_{\gamma \adt},
\ee
and so can be dropped as it annihilates $\mathcal{P}_n$ on its own. The only non-trivial contribution from the first order term in (\ref{k'}) is therefore the third term from (\ref{firstorderterm}),
\be
\sum_{i<j}  \lambda_{j \a} \frac{\del}{\del \mu_j^{\adt}} .
\label{firstcontrib}
\ee

The second order term in (\ref{k'}) acting on momentum twistor space, after using the chain rule (\ref{chain}), becomes
\beqa
&&- \sum_{i<k} \langle i k \rangle \frac{\partial^2}{\partial\lambda_i^{\alpha} \partial\mu_k^{\da}} - \sum_{i<k} \lambda_{k\alpha} \frac{\partial}{\partial\mu_k^{\da}} \\
&& + \sum_{i<k} x_{i\alpha}^{~~\db} \langle i k \rangle \frac{\partial^2}{\partial\mu_i^{\db} \partial\mu_k^{\da}} + \sum_{i<k} \theta_{i\alpha}^A \langle i k \rangle \frac{\partial^2}{\partial\chi_i^{A} \partial\mu_k^{\da}}
\label{ksecond}\\
&& + \sum_{i}\sum_{k,m>i} \tlam_i^{\db} \lambda_{k\alpha}  \langle i m \rangle \frac{\partial^2}{\partial\mu_k^{\db} \partial\mu_m^{\da}}  + \sum_{i}\sum_{k,m>i} \eta_i^A \lambda_{k\alpha}  \langle i m \rangle \frac{\partial^2}{\partial\chi_k^{A} \partial\mu_m^{\da}} .
\label{kthird}
\eeqa
The second term cancels the contribution (\ref{firstcontrib}).
The first term in the third line (\ref{kthird}) contains $\lambda_i \tilde\lambda_i = x_{i,i+1}$. It can be divided into three parts, depending on values of $m$ and $k$ with respect to each other. The first term of (\ref{kthird}) then becomes
\beq
\left(\sum_{i<m<k} + \sum_{i<k<m}\right) x_{i,i+1}^{\db\rho} \lambda_{k\alpha} \lambda_{m\rho} \frac{\partial^2}{\partial\mu_k^{\db} \partial\mu_m^{\da}} + \sum_{i<k=m} x_{i,i+1}^{\db\rho} \lambda_{k\alpha} \lambda_{k\rho} \frac{\partial^2}{\partial\mu_k^{\db} \partial\mu_k^{\da}} .
\label{trisum}
\eeq
The sums over $i$ can now be performed; for instance
\beq
\sum_{i<m<k} x_{i,i+1}^{\db\rho}
= \sum_{m<k} (x_{1}-x_m)^{\db\rho} .
\eeq
The terms proportional to $x_1$ in (\ref{trisum}) sum up together to give
\be
\sum_{k,m} x_1^{\rho \dot\rho} P_{k \rho \dot\alpha} P_{m \alpha \dot\rho} = x_1^{\rho \dot\rho} P_{\rho \dot\alpha} P_{\alpha \dot\rho},
\ee
which can be neglected as $P_{\alpha \dot\alpha} \mathcal{P}_n =0$.
The same procedure applies for the second term in (\ref{kthird}) which yields a terms of the form $\theta_1^{\rho A} Q_{\alpha A} P_{\rho \dot\alpha}$. The remaining terms which depend on $x_i,\theta_i$ combine to give terms which can be written purely in terms of $\lambda_i$, $\mu_i$ and $\chi_i$,
\begin{align}
k'_{\alpha \dot\alpha} \cong &-\sum_{i<k} \biggl[ \langle ik \rangle \frac{\partial^2}{\partial \lambda_i^\alpha \partial \mu_k^{\dot\alpha}} + \lambda_{k\alpha} \mu_i^{\dot\beta} \frac{\partial^2}{\partial \mu_i^{\dot\beta} \partial \mu_k^{\dot\alpha}} + \lambda_{k \alpha} \mu_i^{\dot\beta} \frac{\partial^2}{\partial \mu_k^{\dot\beta} \partial \mu_i^{\dot\alpha}}\biggr] - \sum_k \lambda_{k \alpha} \mu_k^{\dot\beta} \frac{\partial^2}{\partial \mu_k^{\dot\beta} \partial \mu_k^{\dot\alpha}} \notag \\
& - \sum_{i<k} \biggl[ \chi_i^A \lambda_{k \alpha} \frac{\partial^2}{\partial \chi_i^A \partial \mu_k^{\dot\alpha}} - \chi_i^A \lambda_{k \alpha} \frac{\partial^2}{\partial \chi_k^A \partial \mu_i^{\dot\alpha}} \biggr] - \sum_k \chi_k^A \lambda_{k \alpha} \frac{\partial^2}{\partial \chi_k^A \partial \mu_k^{\dot\alpha}}.
\end{align}
Using the helicity condition,
\beq
 \left[ \lambda_{i}^{\alpha}\frac{\partial}{\partial\lambda_i^{\alpha}} + \mu_i^{\da}\frac{\partial}{\partial\mu_i^{\da}} + \chi_i^A\frac{\partial}{\partial\chi_i^{A}} \right] = 0
\eeq
the generator $k'_{\alpha\da}$ can be expressed as the sum of diagonal terms and bilocal terms:
\beqa
k'_{\alpha\da} &=& \sum_k \lambda_{k}^{\rho} \frac{\partial}{\partial\lambda_k^{\rho}} \lambda_{k\alpha} \frac{\partial}{\partial\mu_k^{\da}} \nn \\
&+& \sum_{i<k} \left\{ -\lambda_{i}^{\rho}  \frac{\partial}{\partial\lambda_i^{\alpha}} \lambda_{k \rho} \frac{\partial}{\partial\mu_k^{\da}}
+ \lambda_{i}^{\rho}  \frac{\partial}{\partial\lambda_i^{\rho}} \lambda_{k \alpha} \frac{\partial}{\partial\mu_k^{\da}} - \mu_i^{\dot\rho} \frac{\partial}{\partial\mu_i^{\da}}\lambda_{k \alpha} \frac{\partial}{\partial\mu_k^{\dot\rho}} - \chi_i^A \frac{\partial}{\partial\mu_i^{\da}}\lambda_{k \alpha} \frac{\partial}{\partial\chi_k^{A}}
\right\} .
\eeqa
This actually is the same result, up to an overall normalisation, as the one obtained by inserting  the generators (\ref{dualgen}) in momentum twistor space  into the bilocal formula (\ref{Pone}).
This calculation follows the same lines as above, using the helicity condition, spinor properties and neglecting terms proportional to level-zero generators.

What we have shown is that there are two equivalent ways of looking at the full symmetry algebra of the scattering amplitudes.
The first is as the Yangian of the ordinary superconformal algebra, which if we write it in the twistor representation\footnote{Here the supertwistor variable is $\mathcal{Z}^{\AA} = (\tilde{\mu}^{\alpha}, \tilde\lambda^{\dot\alpha} , \eta^A)$ where $\tilde\mu$ is Fourier conjugate variable of $\lambda$.}, takes the form \cite{Drummond:2009fd},
\begin{align}
j^{\AA}{}_{\BB} &= \sum_i \cZ_i^{\AA} \frac{\del}{\del \cZ_i^{\BB}}, \label{twistorsconf}\\
j^{(1)}{}^{\AA}{}_{\BB} &= \sum_{i<j} (-1)^{\CC}\Bigl[\cZ_i^{\AA} \frac{\del}{\del \cZ_i^{\CC}} \cZ_j^{\CC} \frac{\del}{\del \cZ_j^{\BB}} - (i,j) \Bigr],
\label{twistoryangian}
\end{align}
where both operators are understood to have the supertraces removed. These operators annihilate the amplitude $\mathcal{A}_n$,
\be
j \mathcal{A}_n = j^{(1)} \mathcal{A}_n = 0.
\ee

\begin{figure}
\psfrag{j}[cc][cc]{$j$} \psfrag{J}[cc][cc]{$J$}
\psfrag{j1}[cc][cc]{$j^{(1)}$} \psfrag{J1}[cc][cc]{$J^{(1)}$}
\psfrag{pq}[cc][cc]{\!\!\!\!\!\!\!\!\!\!\!$p,q$}\psfrag{PQ}[cc][cc]{\,\,\,\,$P,Q$}
\psfrag{KS}[cc][cc]{\!\!\!\!\!\!\!\!\!\!\!$K,S$}\psfrag{ks}[cc][cc]{\,\,\,\,$k,s$}
\psfrag{rR}[cc][cc]{}
\psfrag{Tduality}[cc][cc]{T-duality}
\centerline{{\epsfysize9cm \epsfbox{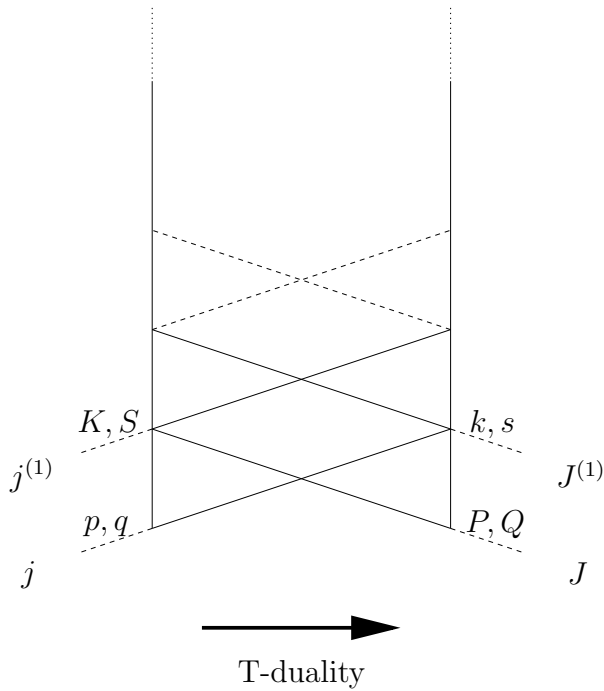}}} \caption[]{\small The tower of symmetries acting on scattering amplitudes in $\cN=4$ super Yang-Mills theory. The original superconformal charges are denoted by $j$ and the dual ones by $J$. Each can be thought of as the level-zero part of the Yangian $Y(psu(2,2|4))$. The dual superconformal charges $K$ and $S$ form part of the level-one $j^{(1)}$ while the original superconformal charges $k$ and $s$ form part of the level one charges $J^{(1)}$. In each representation the `negative' level ($P$ and $Q$ or $p$ and $q$) is trivialised. T-duality maps $j$ to $J$ and $j^{(1)}$ to $J^{(1)}$.}
\label{Fig:aux}
\end{figure}

The second way of writing the symmetry is as the Yangian of the dual superconformal algebra, which, written in the momentum twistor representation, takes an identical form up to the change from twistors to momentum twistors,
\begin{align}
J^{\AA}{}_{\BB} &= \sum_i \cW_i^{\AA} \frac{\del}{\del \cW_i^{\BB}},\label{momtwistordsconf}\\
J^{(1)}{}^{\AA}{}_{\BB} &= \sum_{i<j} (-1)^{\CC}\Bigl[\cW_i^{\AA} \frac{\del}{\del \cW_i^{\CC}} \cW_j^{\CC} \frac{\del}{\del \cW_j^{\BB}} - (i,j) \Bigr].
\label{momtwistoryangian}
\end{align}
These operators annihilate the amplitude with the MHV amplitude factored out,
\be
\mathcal{A}_{n} = \mathcal{A}_n^{\rm MHV} \mathcal{P}_n, \qquad J \mathcal{P}_n = J^{(1)} \mathcal{P}_n = 0.
\ee
The picture we find is very natural from the point of view of T-duality in the $AdS$ sigma model. In \cite{Berkovits:2008ic,Beisert:2008iq,Beisert:2009cs} it was shown that the supersymmetric $AdS_5 \times S^5$ background maps into itself as does the infinite tower of conserved charges associated with the integrability of the sigma model \cite{Bena:2003wd}.

\section{Grassmannian formulas}
\label{sect-grassmannians}

The feature that we have just seen is also natural from another perspective. Recently some remarkable formulas have been proposed as a way of computing all the leading singularities
of  $\cN=4$ super Yang-Mills amplitudes. These formulas take the form of an integral over the Grassmannian $G(k,n)$ of certain superconformally invariant delta functions.
In the original proposal of \cite{ArkaniHamed:2009dn}, the integral takes the following form,
\be
\mathcal{L}_{\rm ACCK} = \int \frac{\prod_{a,i} dc_{ai}}{(1\ldots k)(2\ldots k+1)\ldots(n \ldots n+k-1)} \prod_{a=1}^k \delta^{4|4}\Bigl(\sum_{i=1}^n c_{ai} \cZ_i\Bigr).
\label{ACCK}
\ee
Here one considers a $(k \times n)$ matrix of complex parameters $c_{ai}$ which are integrated over certain contours which have to be specified\footnote{Note that here and in the next section we use the indices $a,b=1,\ldots,k$ to denote the rows of the $k\times n$ matrix, rather than adjoint indices of $psu(2,2|4)$ as in the previous sections. We hope that the context will be sufficient to avoid confusion.}. The delta functions are manifestly invariant under ordinary superconformal symmetry (in its twistor representation (\ref{twistorsconf})).

The denominator consists of the cyclic product of determinants of $(k \times k)$ submatrices (or minors) of the large $(k \times n)$ matrix of the $c_{ai}$. For example the notation $(1 \ldots k)$ means the minor made from the first $k$ columns of the full matrix of $c_{ai}$.
As described in \cite{ArkaniHamed:2009dn}, the integral measure should be carefully defined in (\ref{ACCK}), taking into account the fact that the integral possesses a $GL(k)$ gauge symmetry. One can do this by fixing a gauge such that $k$ columns of the matrix of the $c_{ai}$ become the $(k \times k)$ identity matrix. Then one integrates over the unfixed $c_{ai}$ in two steps. First one uses the delta functions of the bosonic variables to determine as many of the $c_{ai}$ as possible and reconstruct the momentum conserving delta function. Then one chooses a specific contour of integration for the remaining $c_{ai}$. Different choices of contour lead to different expressions but remarkably each expression so obtained seems to have a role to play in the amplitude $\mathcal{A}_n$ as an integral coefficient in the expansion (\ref{integralexpansion}). One can obtain coefficients which appear in the tree-level amplitude as well as one-loop and even higher-loop integral coefficients in this way.
There are $4k$ Grassmann delta functions in the original integral and so these expressions appear in ${\rm N}^{k-2}{\rm MHV}$ amplitudes.

A very similar formula to (\ref{ACCK}) was proposed in \cite{Mason:2009qx}. The difference is that it is written in terms of momentum twistors, instead of twistors and therefore it is the dual superconformal symmetry which is manifest,
\be
\mathcal{L}_{\rm MS} = \int \frac{\prod_{a,i} dt_{ai}}{(1\ldots k)(2\ldots k+1)\ldots(n \ldots n+k-1)} \prod_{a=1}^k \delta^{4|4}\Bigl(\sum_{i=1}^n t_{ai} \cW_i\Bigr).
\label{MS}
\ee
The structure of the formula is identical to (\ref{ACCK}), with the integration variables called $t_{ai}$ forming a $(k \times n)$ matrix. This time the formula generates contributions to $\mathcal{P}_n$ (instead of $\mathcal{A}_n$), in other words it produces the same quantities (but written in different variables) as (\ref{ACCK}) but with the MHV tree-level amplitude factored out. Thus the $4k$ Grassmann delta functions mean that this formula generates contributions to ${\rm N}^k{\rm MHV}$ amplitudes.

In fact it has been shown that the two formulas are related by change of variables from one to the other \cite{ArkaniHamed:2009vw}. This shows indirectly that both formulas actually possess the non-manifest superconformal symmetries, the dual superconformal symmetry for (\ref{ACCK}) and the ordinary superconformal for (\ref{MS}). This suggests that the Grassmannian integral formula should be interpreted as the general form of an invariant under the full Yangian symmetry (in either version as they are simply related by a change of variables). Here we recall that the leading singularities are obtained from products of tree-level amplitudes. Hence we expect them to be invariant under the action of the Yangian generators (\ref{twistorsconf},\ref{twistoryangian}) or equivalently (\ref{momtwistordsconf},\ref{momtwistoryangian}) for generic kinematical configurations. There will be contact-type anomalies for singular kinematical configurations \cite{Bargheer:2009qu,Korchemsky:2009hm,Sever:2009aa}. As we are considering the generic case, we do not deform the free representations  (\ref{twistorsconf},\ref{twistoryangian}) and (\ref{momtwistordsconf},\ref{momtwistoryangian}), as is done in \cite{Bargheer:2009qu,Sever:2009aa}.


\section{Yangian invariance of the Grassmannian formulas}
\label{sect-inv}

We would like to show that the Yangian generators (\ref{twistorsconf},\ref{twistoryangian}) and (\ref{momtwistordsconf},\ref{momtwistoryangian}) provide a natural and direct way to show the non-manifest invariance of each of the Grassmannian formulas. One reason for wanting to show invariance directly is to develop a method which will might allow a proof that the Grassmannian integral is in fact the most general form of an invariant under the Yangian symmetry.
As we have seen the Yangian symmetry looks the same in either twistor or momentum twistor versions so it will not matter (at least formally) which version we consider here. To be concrete we will take the momentum twistor representations of the Yangian symmetry (\ref{momtwistordsconf},\ref{momtwistoryangian}) and the Grassmannian formula (\ref{MS}). This will permit us to use a manifestly $psu(2,2|4)$ invariant language without having to worry about taking a Fourier transform which is justified only in (2,2) signature. The calculation we will perform is equivalent to directly showing the original superconformal invariance of (\ref{MS}).

We will first work with the formal integral in which no gauge-fixing has been performed and keep the full (though ill-defined) set of integrations over all of the $t_{ai}$ parameters. This will reveal some general features that will allow us to perform a more honest calculation where the integral is gauge-fixed and well-defined.

So we will consider the formal expression
\be
\mathcal{L}_{n,k} = \int \frac{\prod_{a,i} dt_{ai}}{\MM_1 \ldots \MM_n} \prod_{a=1}^k \delta_a.
\label{formalL}
\ee
Here $\MM_p$ stands for the consecutive $k \times k$ minor made from the columns $
p,\ldots,p+k-1$ of the $k \times n$ matrix of the $t_{ai}$,
\beq
\MM_{p} \equiv (p~p+1~p+2\ldots p+k-1)
\eeq 	
and we have introduced the shorthand notation for the delta functions from (\ref{MS}),
\be
\delta_a = \delta^{4|4}\Bigl(\sum_{i=1}^n t_{ai} \cW_i\Bigr).
\label{deltas}
\ee
The expression (\ref{formalL}) is manifestly invariant under the level-zero generators (\ref{momtwistordsconf}) being made of the dual superconformally invariant delta functions (\ref{deltas}). To show the Yangian symmetry we need to act on it with the level-one generator (\ref{momtwistoryangian}). In fact we can drop the antisymmetrisation on the indices $i$ and $j$ and consider instead the operator (as usual understood to be supertraceless),
\be
\sum_{i<j}(-1)^{\CC}\biggl[ \cW_i^{\AA} \frac{\partial}{\partial \cW_i^{\CC}} \cW_j^{\CC} \frac{\partial}{\partial \cW_j^{\BB}} \biggr]. \label{J1mod}
\ee
This is because we can write the operator in (\ref{momtwistoryangian}) as
\begin{align}
J^{(1)}{}^{\AA}{}_{\BB} &= \biggl(\sum_{i<j} - \sum_{j<i}\biggr) (-1)^{\CC}  \cW_i^{\AA} \frac{\partial}{\partial \cW_i^{\CC}} \cW_j^{\CC} \frac{\partial}{\partial \cW_j^{\BB}} \notag \\
&= \biggl(2\sum_{i<j} - \sum_{i,j} + \sum_{i=j} \biggr) (-1)^{\CC}  \cW_i^{\AA} \frac{\partial}{\partial \cW_i^{\CC}} \cW_j^{\CC} \frac{\partial}{\partial \cW_j^{\BB}}.
\end{align}
The second and third summations annihilate the delta functions on their own as they can be shown to be proportional to level-zero generators. The first summation gives the operator (\ref{J1mod}) up to a factor of two.

We can rewrite each term in the operator (\ref{J1mod}) in the following way (recall $i\neq j$),
\begin{align}
(-1)^C \cW_i^{\AA} \frac{\del}{\del \cW_{i}^{\CC}} \cW_j^{\CC} \frac{\del}{\del \cW_j^{\BB}} &= \cW_i^{\AA} \cW_j^{\CC} \frac{\del}{\del \cW_i^{\CC}} \frac{\del}{\del \cW_j^{\BB}} \\
&= (-1)^{\BB \CC} \cW_i^{\AA} \cW_j^{\CC} \frac{\del}{\del \cW_j^{\BB}} \frac{\del}{\del \cW_i^{\CC}} \\
&= \cW_i^{\AA} \Bigl( \frac{\del}{\del \cW_j^{\BB}} \cW_j^{\CC} - \delta^{\CC}_{\BB} \Bigr) \frac{\del}{\del \cW_i^{\CC}}\\
&= \cW_i^{\AA} \frac{\del}{\del \cW_j^{\BB}} \cW_j^{\CC} \frac{\del}{\del \cW_i^{\CC}} - \cW_i^{\AA} \frac{\del}{\del \cW_i^{\BB}}.
\label{orderedj1}
\end{align}
Now the first term of (\ref{orderedj1}) contains the operator
\be
\cW_j^{\CC} \frac{\del}{\del \cW_i^{\CC}} \label{glnW}
\ee
which acts as a $gl(n)$ transformation on the $\cW_i$. The delta functions are $gl(n)$ invariant if we transform the $t_{ai}$ in the opposite way. Hence on the delta functions we can replace the operator (\ref{glnW}) with
\be
\OO_{ij} = \sum_{a=1}^k  t_{a i} \frac{\partial}{\partial t_{a j}}.
\label{Oij}
\ee
In other words the action of the Yangian generator induces a particular compensating $gl(n)$ transformation of the $t_{ai}$ variables.

To summarise, we have found that the action of the level-one operator $J^{(1)}{}^{\AA}{}_{\BB}$ on the Grassmannian formula $\mathcal{L}_{n,k}$ (\ref{formalL}) can be written as
\be
\tfrac{1}{2} J^{(1)}{}^{\AA}{}_{\BB} \mathcal{L}_{n,k} =
 \int  \frac{\prod_{a,m} dt_{am}}{\MM_1 \MM_2\ldots \MM_n}  \sum_{i<j}\biggl[
\mathcal{O}_{ij} \cW_i^{\AA} \frac{\del}{\del \cW_j^{\BB}} - \cW_i^{\AA} \frac{\del}{\del \cW_i^{\BB}} \biggr] \prod_{a=1}^k \delta_a ~,
\label{invariance}
\ee
The $\cW$-derivatives in (\ref{invariance}) act on each $\delta$-function in turn, giving a sum of similar contributions,
\be
\frac{\partial}{\partial \cW_i^{\BB}} \prod_{a=1}^k \delta_a = \sum_{b=1}^{k} t_{bi} \bigl(\partial_{\BB} \delta_b\bigr) \prod_{a\neq b} \delta_a.
\ee
Using this on both terms in the square brackets in (\ref{invariance}), the level-one variation becomes
\be
\sum_b  \int  \frac{\prod_{a,m} dt_{am}}{\MM_1 \MM_2\ldots \MM_n}  [\mathcal{O}^{\AA}_b - \mathcal{V}^{\AA}_b] \bigl( \partial_{\BB} \delta_b \bigr) \prod_{a\neq b} \delta_a.
\label{invOAb}
\ee
where the first-order operator $\mathcal{O}^{\AA}_b$ (which generates a particular triangular $gl(n)$ transformation by commutation) is given by
\be
\mathcal{O}^{\AA}_b = \sum_{i<j}\cW_i^{\AA} \mathcal{O}_{ij} t_{bj}
\label{OAb}
\ee
and $\mathcal{V}^{\AA}_b$ is simply given by
\be
\mathcal{V}^{\AA}_b = \sum_{i<j} \cW^{\AA}_i t_{bi}.
\ee

The idea now is to commute the operator $\mathcal{O}^{\AA}_b$ back past the minors in the denominator.
When the operator reaches the measure $\prod dt_{am}$, it will be a total derivative (recall $i\neq j$ in the sum) and (at least formally) can be neglected. In commuting the operator $\mathcal{O}^{\AA}_b$ past the minors we will pick up a sum of terms as they are not invariant,
\be
\biggl[\frac{1}{\MM_1 \ldots \MM_n} , \mathcal{O}^{\AA}_b\biggr] \neq 0. \label{commutator}
\ee
In fact this variation will precisely cancel the $\mathcal{V}^{\AA}_b$ term in (\ref{invOAb}). The essential reason that the commutator is non-vanishing is that the minors are {\it not} invariant under $gl(n)$ transformations. Indeed the action of the $gl(n)$ generator $\mathcal{O}_{ij}$ on a general minor of the form $\mathcal{M}_p$ is simply to replace the entry $j$ in $\MM_p$ by $i$ if $j$ is present,
\be
\mathcal{O}_{ij} \mathcal{M}_p = \sum_{a=1}^k t_{ai} \frac{\partial}{\partial t_{aj}} \MM_p =  \mathcal{M}_p^{j \ra i} \equiv (p \ldots j-1 \, i \, j+1 \ldots p+k-1).
\label{OijMM}
\ee
and is vanishing if the entry $j$ is not present. Obviously the result (\ref{OijMM}) vanishes if $i$ is already present as another entry in $\MM_p$ due to antisymmetry.

Using (\ref{OijMM}) a short calculation (which we present in appendix \ref{app-comm}) shows that under the triangular $gl(n)$ transformation generated by $\mathcal{O}^{\AA}_b$ we have
\be
[\mathcal{O}^{\AA}_b, \MM_p] = \biggl( \sum_{i=1}^{p-1} \cW^{\AA}_i t_{bi}\biggr) \, \MM_p.
\label{Mvar}
\ee
In other words, the consecutive minor $\MM_p$ transforms into itself up to a factor. Note the privileged role of the consecutive minors as opposed to general minors $(i_1 \ldots i_k)$ which do not transform covariantly.
It is now simple to compute the commutator we need from (\ref{commutator}) and we find
\be
\biggl[\frac{1}{\MM_1 \ldots \MM_n} , \mathcal{O}^{\AA}_b\biggr] = \frac{1}{\MM_1 \ldots \MM_n} \sum_{p=1}^n \frac{[\mathcal{O}^{\AA}_b, \MM_p]}{\MM_p}  =  \frac{1}{\MM_1 \ldots \MM_n} \sum_{i<p} \cW^{\AA}_i t_{bi} = \frac{\mathcal{V}^{\AA}_b}{\MM_1 \ldots \MM_n}.
\ee
As anticipated this is precisely what is needed to cancel the $\mathcal{V}^{\AA}_b$ term from (\ref{invOAb}).

Thus we have shown that the only contribution to the level-one variation (\ref{invariance}) is the total derivative term where the $gl(n)$ operator $\OO_{ij}$ reaches the integration measure,
\be
\sum_b \sum_{i<j} \int \prod_{a,m} dt_{am} \OO_{ij}  \biggl[ \cW_i^{\AA} t_{bj} \frac{1}{\MM_1 \MM_2\ldots \MM_n} \bigl( \partial_{\BB} \delta_b \bigr) \prod_{a\neq b} \delta_a \biggr].
\ee
Formally this term can be neglected as it is an integral of a total derivative. Therefore, $\mathcal{L}_{n,k}$ is formally invariant under the Yangian symmetry for generic $n$ and $k$ if the integration is performed over any closed contour. To state what we have shown in a coordinate invariant way, the form being integrated varies up to a total derivative
\be
J^{(1)}{}^{\AA}{}_{\BB} K = d \Omega^{\AA}{}_{\BB}.
\label{exactvar}
\ee
Therefore, for any closed contour the variation will integrate to zero. Of course if the integration region has boundaries then the total derivatives can contribute boundary terms and hence imply a breaking of the symmetry.
The reason that what we have shown is only formally a proof of invariance is that the integration over all of the $t_{ai}$ is not well-defined.

We would now like to work with a well-defined finite integral and show Yangian invariance. The problem with the formal integral (\ref{formalL}) is the $gl(k)$ gauge redundancy. There are two options for rendering this well-defined. We could work gauge-invariantly and use the $(n-k)\times k$-dimensional gauge-invariant measure given by Mason and Skinner \cite{Mason:2009qx} and show that it is invariant under the effective transformation of the $t_{ai}$ generated by $\mathcal{O}^{\AA}_b$. Alternatively we could fix this measure to a convenient gauge and show invariance directly on the gauge-fixed integral. Since the initial integral is gauge-invariant this is sufficient to show invariance in any gauge.
The second option proves to be remarkably simple so we will pursue this approach. The gauge we will choose is the one where we fix the first $k$ columns of the matrix $t_{ai}$ to be the identity matrix,
\beq
\left(
\begin{array}{ccc|ccc}
{} & {} & {}  & t_{1 k+1} & \ldots & t_{1n}\\
{} & \mathds{1}_{k \times k}  & {} & {\vdots} & {} & {\vdots}\\
{} & {} & {}  & {t_{k k+1}} & {\ldots} & {t_{kn}}\end{array}
\right).
\eeq
The integration is now over the remaining $(n-k)\times k$  variables,
\be
\prod_{a,m} dt_{am} = \prod_{a=1}^k \prod_{m=k+1}^n dt_{am}.
\ee
Since some of the $t_{ai}$ are now 0 or  1 the integrand is simplified. In particular the delta functions become
\be
\delta_a = \delta^{4|4}\Bigl( \cW_a + \sum_{l=k+1}^n t_{al} \cW_l \Bigr).
\ee

The only difference in calculating the level-one variation of this gauge-fixed integral from what we did before is the step from (\ref{glnW}) to (\ref{Oij}) where we replaced the operator
\be
\cW^{\CC}_j \frac{\partial}{\partial \cW^{\CC}_i}   \longrightarrow \mathcal{O}_{ij} = \sum_a t_{ai} \frac{\partial}{\partial t_{aj}}.
\ee
This is still fine if $j>k$ but if $j\leq k$ then we run into the gauge-fixed parts of the delta functions and we must treat the operator differently. In fact we can rewrite it in the following way (recalling that $i<j\leq k$),
\begin{align}
\cW_j^{\CC}\frac{\partial}{\partial \cW_i^{\CC}} &\prod_a \delta_a
= \cW_j^{\CC}\biggl[\frac{\partial}{\partial \cW_i^{\CC}} \delta_i \biggr] \prod_{a\neq i} \delta_a,
\end{align}
where we used the fact that $i<j\leq k$ and therefore the  variable $\cW_i$ is present only in one specific  delta function. This result can be rewritten as a function of $\cW_r$, with $r>k$, by means of the constraint in  the delta function:
\begin{align}
\biggl[-\sum_{r=k+1}^n t_{jr} \cW^{\CC}_r \biggr]\biggl[ \frac{\partial}{\partial \cW_i^{\CC}}  \delta_i\biggr] \prod_{a\neq i} \delta_a ~.
 \end{align}
For each term in the sum over $r$ we can exchange the $\cW_i$ derivative for a $\cW_r$ derivative as follows,
\begin{align}
\biggl[-\sum_{r=k+1}^n t_{jr} \cW^{\CC}_r
 \frac{1}{t_{ir}} \frac{\partial}{\partial \cW_r^{\CC}}  \delta_i \biggr] \prod_{a\neq i} \delta_a ~.
 \end{align}
Since the resulting operator generates a scaling of $\cW_r$, on $\delta_i$ we can replace it with a scaling of $t_{ir}$ instead,
 \beq
 \cW^{\CC}_r  \frac{\partial}{\partial \cW_r^{\CC}}   \delta_i =   t_{ir} \frac{\partial}{\partial t_{ir}} \delta_i ~
 \eeq
and we arrive finally at
\begin{align}
\cW_j^{\CC} \frac{\partial}{\partial \cW_i^{\CC}} \prod_{a} \delta_a
 = - \mathcal{U}_{ij} \prod_a \delta_a
\end{align}
where we have defined
\be
\UU_{ij} = \sum_{r=k+1}^n  t_{j r} \frac{\partial}{\partial t_{i r}}, \qquad 1\leq i < j \leq k.
\label{Uij}
\ee

Note that in $\UU_{ij}$, the labels $i$ and $j$ denote the row indices of the matrix of $t$'s, in contrast to the labels of $\mathcal{O}_{ij}$ where they are column indices.  Indeed the operator $\mathcal{U}_{ij}$ acts as a $gl(k)$ rotation on the rows of the non-gauge-fixed part of this matrix. Thus it acts on minors by replacing the $i$-th row by the $j$-th one on the non-gauge-fixed part of the matrix of $t$'s (recall that $r > k$ in the sum). Therefore
\be
\UU_{ij} \MM_p = 0~~ \mathrm{if} ~~ k<p\leq(n-k)
\ee
as $\UU_{ij} \MM_p$ is the determinant of a matrix with two equal rows. For $(n-k)<p\leq n$ the result is also vanishing. The only non-vanishing contribution is given when $\UU_{ij}$ acts on a minor $\MM_p$ with $1<p\leq k$. As we explain in appendix \ref{app:derivationU}, after a careful study one can convince oneself  that its action is equivalent, up to a sign, to replacing the $j$-th column by the $i$-th one. Therefore
\be
\UU_{ij} \MM_p = - \MM_p^{j\rightarrow i}~~ \mathrm{if} ~~ 1<p\leq k ~~, ~~i<j\leq k
\label{Uijtransform}
\ee
which is exactly the same result for $\OO_{ij}$ (\ref{OijMM}), apart from a sign. We can therefore unify the two operators into a single operator $\NN_{ij}$ valid for all values of $j$,
\beq
\NN_{ij} = \left( -\UU_{ij}, \OO_{ij}\right).
\eeq
and then define
\be
\mathcal{N}^{\AA}_b = \sum_{i<j}\cW_i^{\AA} \mathcal{N}_{ij} t_{bj}~.
\ee
The operator $\NN^{\AA}_b$ is the gauge-fixed version of $\OO^{\AA}_b$ from (\ref{OAb}).
Following the same steps as in the gauge-invariant case, the level-one variation becomes
\be
\sum_b  \int  \frac{\prod_{a,m} dt_{am}}{\MM_1 \MM_2\ldots \MM_n}  [\mathcal{N}^A_b - \mathcal{V}^A_b] \bigl( \partial_B \delta_b \bigr) \prod_{a\neq b} \delta_a.
\label{invNAb}
\ee
As before, one can commute the operator $\mathcal{N}^{\AA}_b$ back past the minors in the denominator. The steps are identical to the gauge-invariant case we discussed previously. In particular, due to (\ref{Uijtransform}), the minors transform as before
\be
[\mathcal{N}^{\AA}_b, \MM_p] = \biggl( \sum_{i=1}^{p-1} \cW^{\AA}_i t_{bi}\biggr) \, \MM_p.
\label{MvarN}
\ee
The remaining term is then a true total derivative,
\be
\sum_b \sum_{i<j} \int \Bigl( \prod_{a,m} dt_{am} \Bigr) \NN_{ij} \biggl[ \cW_i^{\AA} t_{bj}   \frac{1}{\MM_1 \MM_2\ldots \MM_n} \bigl( \partial_{\BB} \delta_b \bigr) \prod_{a\neq b} \delta_a \biggr],
\ee
i.e. we have shown that (\ref{exactvar}) holds.
This completes the direct proof of the Yangian invariance of the Grassmannian formulas.

\section{Conclusions}

In this paper we have considered the Yangian symmetry of scattering amplitudes in $\NN = 4$ SYM theory. In \cite{Drummond:2009fd} it was shown that the ordinary superconformal symmetry forms the level-zero subalgebra of a Yangian algebra with the dual superconformal symmetry providing part of the level-one generators. The remaining generators are obtained from these by commutation.  Here we have shown that there is a `T-dual' version, where the roles of the original and dual superconformal symmetries are interchanged. In this case, the Yangian generators annihilate the amplitude with the MHV part factored out, rather than the whole amplitude. The momentum twistors of \cite{Hodges:2009hk} played an important role in this analysis, indeed the representation of the T-dual version of the Yangian in terms of the  momentum twistors is identical to that of the original version in terms of the usual twistors.

The T-duality structure is reflected in recently proposed Grassmannian formulas which reproduce
{leading singularities}
of scattering amplitudes. The first proposal \cite{ArkaniHamed:2009dn}, formulated in twistor space, is manifestly invariant under ordinary superconformal symmetry, while the formulation in momentum twistor space \cite{Mason:2009qx} is invariant under dual superconformal symmetry. The two formulas are related by a change of variables \cite{ArkaniHamed:2009vw} which shows indirectly that they both have the ordinary and dual superconformal symmetries, and that the objects they produce are Yangian invariants. It is tempting to regard the Grassmannian formula as the most general form of an invariant under the Yangian symmetry. Then the fact that the two versions have precisely the same structure (one simply exchanges twistors for momentum twistors) is a natural expression of the T-duality structure of the Yangian itself.
In this paper, we have directly proved the Yangian invariance of these Grassmannian formulas by using the explicit expression of the level-one generators. In our calculation, to be concrete, we used the momentum twistor version but we could equally well have used the twistor version as the two formulas are identical in structure. In the proof we saw explicitly the role of the $gl(n)$ invariance of the delta functions and the $gl(k)$ gauge symmetry.

We think that one of the main issues to address is to demonstrate that the most general invariant under the Yangian symmetry takes exactly the form of the Grassmannian integral. The methods we have developed in this paper may turn out to be very useful in this respect. Further interesting questions remain open in this context. For instance, the contribution of the holomorphic anomaly to these formulas on singular kinematical configurations and the extension of the Yangian symmetry to loop level.


\section*{Acknowledgements}
We would like to thank Emery Sokatchev for many interesting discussions. This research was supported in part by the French Agence Nationale de la Recherche under grant ANR-06-BLAN-0142.

\appendix

\setcounter{section}{0} \setcounter{equation}{0}
\renewcommand{\theequation}{\Alph{section}.\arabic{equation}}

\section{Formulae for both superconformal algebras}
We
begin by listing the commutation relations of the algebra $u(2,2|4)$. The Lorentz generators
$\mathbb{M}_{\a \b}$, $\overline{\mathbb{M}}_{\adt \bdt}$ and the $su(4)$ generators
$\mathbb{R}^{A}{}_{B}$ act canonically on the remaining generators carrying Lorentz or $su(4)$
indices. The dilatation $\mathbb{D}$ and hypercharge $\mathbb{B}$ act via
\be
[\mathbb{D},\mathbb{J}] = {\rm dim}(\mathbb{J})\, \mathbb{J}, \qquad [\mathbb{B},\mathbb{J}] = {\rm
hyp}(\mathbb{J})\, \mathbb{J}.
\ee
The non-zero dimensions and hypercharges of the various generators are
\begin{align} \notag
& {\rm dim}(\mathbb{P})=1, \qqquad {\rm dim}(\mathbb{Q}) = {\rm dim}(\overline{\mathbb{Q}}) =
\tfrac{1}{2},\qquad {\rm dim}(\mathbb{S}) = {\rm dim}(\overline{\mathbb{S}}) = -\tfrac{1}{2}
\\
&{\rm dim}(\mathbb{K})=-1,\qquad {\rm hyp}(\mathbb{Q}) = {\rm hyp}(\overline{\mathbb{S}}) =
\tfrac{1}{2}, \qquad~ {\rm hyp}(\overline{\mathbb{Q}}) = {\rm hyp}(\mathbb{S}) = - \tfrac{1}{2}.
\end{align}
The remaining non-trivial commutation relations are,
\begin{align} \notag
& \{\mathbb{Q}_{\a A},\overline{\mathbb{Q}}_{\adt}^B\}  =  \delta_A^B \mathbb{P}_{\a \adt},
   \qquad \{\mathbb{S}_{\a}^A,\overline{\mathbb{S}}_{\adt B} \} = \delta_B^A \mathbb{K}_{\a \adt},
\\ \notag
& {}[\mathbb{P}_{\a \adt},\mathbb{S}^{\b A}] = \delta_{\a}^{\b} \overline{\mathbb{Q}}_{\adt}^A,
 \qqquad [\mathbb{K}_{\a \adt},\mathbb{Q}^{\b}_{A}] = \delta_{\a}^{\b}
   \overline{\mathbb{S}}_{\adt A},
\\ \notag
& {}[\mathbb{P}_{\a \adt},\overline{\mathbb{S}}^{\bdt}_{A}]  =  \delta^{\bdt}_{\adt} \mathbb{Q}_{\a A},
\qqquad [\mathbb{K}_{\a \adt}, \overline{\mathbb{Q}}^{\bdt A}]  =  \delta_{\adt}^{\bdt} \mathbb{S}_{\a}^{A},
\\ \notag
& [\mathbb{K}_{\a \adt},\mathbb{P}^{\b \bdt}] = \delta_\a^\b \delta_\adt^\bdt \mathbb{D} +
\mathbb{M}_{\a}{}^{\b}
 \delta_\adt^\bdt + \overline{\mathbb{M}}_{\adt}{}^{\bdt} \delta_\a^\b,
\\ \notag
& \{\mathbb{Q}^{\a}_{A},\mathbb{S}_\b^B\} =  \mathbb{M}^{\a}{}_{\b}
\delta_A^B + \delta^{\a}_{\b} \mathbb{R}^{B}{}_{A} + \tfrac{1}{2}\delta^{\a}_{\b} \delta_A^B (\mathbb{D}+\mathbb{C}),
\\
& \{\overline{\mathbb{Q}}^{\adt A},\overline{\mathbb{S}}_{\bdt B}\} = \overline{\mathbb{M}}^{\adt}{}_{\bdt} \delta_B^A  - \delta^{\adt}_{\bdt} \mathbb{R}^{A}{}_{B} + \tfrac{1}{2} \delta^{\adt}_{\bdt}\delta_B^A
(\mathbb{D}-\mathbb{C}).
\label{comm-rel}
\end{align}
Note that in writing the algebra relations we are obliged to choose the $su(4)$ chirality of the odd generators. The relations above are valid directly for the dual superconformal generators. For the conventional realisation of the algebra, one should simply swap all $su(4)$ chiralities appearing in the commutation relations.
We now give the generators in both the conventional and dual representations of the superconformal
algebra. We will use the following shorthand notation:
\begin{align}\label{shortderiv}
\partial_{i \alpha \dot{\alpha}} = \frac{\partial}{\partial
x_i^{\alpha \dot{\alpha}}}, \qquad \partial_{i \alpha A} = \frac{\partial}{\partial \theta_i^{\alpha
A}}, \qquad \partial_{i \alpha} = \frac{\partial}{\partial \lambda_i^{\alpha}}\,, \qquad
\partial_{i \dot{\alpha}} = \frac{\partial}{\partial
    \tilde{\lambda}_i^{\dot{\alpha}}}\,, \qquad
\partial_{i A} = \frac{\partial}{\partial \eta_i^A}\,.
\end{align}
We first give the generators of the conventional superconformal symmetry, using lower case
characters to distinguish these generators from the dual superconformal generators which follow
afterwards.
\begin{align}
& p^{\dot{\alpha}\alpha }  =  \sum_i \tilde{\lambda}_i^{\dot{\alpha}}\lambda_i^{\alpha} \,, & &
k_{\alpha \dot{\alpha}} = \sum_i \partial_{i \alpha} \partial_{i \dot{\alpha}} \,,\notag\\
&\overline{m}_{\dot{\alpha} \dot{\beta}} = \sum_i \tilde{\lambda}_{i (\dot{\alpha}} \partial_{i
\dot{\beta} )}, & & m_{\alpha \beta} = \sum_i \lambda_{i (\alpha} \partial_{i \beta )}
\,,\notag\\
& d =  \sum_i [\tfrac{1}{2}\lambda_i^{\alpha} \partial_{i \alpha} +\tfrac{1}{2}
\tilde{\lambda}_i^{\dot{\alpha}} \partial_{i
    \dot{\alpha}} +1], & & r^{A}{}_{B} = \sum_i [-\eta_i^A \partial_{i B} + \tfrac{1}{4}\delta^A_B \eta_i^C \partial_{i C}]\,,\notag\\
&q^{\alpha A} =  \sum_i \lambda_i^{\alpha} \eta_i^A \,, &&   \bar{q}^{\dot\alpha}_A
= \sum_i \tilde\lambda_i^\adt \partial_{i A} \,, \notag\\
& s_{\alpha A} =  \sum_i \partial_{i \alpha} \partial_{i A}, & &
\bar{s}_{\dot\alpha}^A = \sum_i \eta_i^A \partial_{i \dot\alpha}\,,\notag\\
&c = \sum_i [1 + \tfrac{1}{2} \lambda_i^{\a} \partial_{i \a} - \tfrac{1}{2} \tilde\lambda^{\adt}_i \partial_{i \adt} - \tfrac{1}{2} \eta^A_i \partial_{iA} ]\,.
\end{align}
We can construct the generators of dual superconformal transformations by starting with the standard
chiral representation and extending the generators so that they commute with the constraints,
\be
(x_i-x_{i+1})_{\a \dot\alpha}  - \lambda_{i\, \a}\, \tilde{\lambda}_{i\, \dot\alpha} = 0\,, \qquad (\theta_i - \theta_{i+1})_\alpha^A - \lambda_{i \alpha} \eta_i^A = 0\,.
\ee
By construction they preserve the surface defined by these constraints, which is where the amplitude
has support. The generators are
\begin{align}
P_{\alpha \dot{\alpha}}&= \sum_i \partial_{i \alpha \dot{\alpha}}\,, \qquad Q_{\alpha A} = \sum_i \partial_{i \alpha A}\,, \qquad
\overline{Q}_{\dot{\alpha}}^A = \sum_i [\theta_i^{\alpha A}
  \partial_{i \alpha \dot{\alpha}} + \eta_i^A \partial_{i \dot{\alpha}}], \notag\\
M_{\alpha \beta} &= \sum_i[x_{i ( \alpha}{}^{\dot{\alpha}}
  \partial_{i \beta ) \dot{\alpha}} + \theta_{i (\alpha}^A \partial_{i
  \beta) A} + \lambda_{i (\alpha} \partial_{i \beta)}]\,, \qquad
\overline{M}_{\dot{\alpha} \dot{\beta}} = \sum_i [x_{i
    (\dot{\alpha}}{}^{\alpha} \partial_{i \dot{\beta} ) \alpha} +
  \tilde{\lambda}_{i(\dot{\alpha}} \partial_{i \dot{\beta})}]\,,\notag\\
R^{A}{}_{B} &= \sum_i [\theta_i^{\alpha A} \partial_{i \alpha B} +
  \eta_i^A \partial_{i B} - \tfrac{1}{4} \delta^A_B \theta_i^{\alpha
    C} \partial_{i \alpha C} - \tfrac{1}{4}\delta^A_B \eta_i^C \partial_{i C}
]\,,\notag\\
D &= \sum_i [-x_i^{\dot{\alpha}\alpha}\partial_{i \alpha \dot{\alpha}} -
  \tfrac{1}{2} \theta_i^{\alpha A} \partial_{i \alpha A} -
  \tfrac{1}{2} \lambda_i^{\alpha} \partial_{i \alpha} -\tfrac{1}{2}
  \tilde{\lambda}_i^{\dot{\alpha}} \partial_{i \dot{\alpha}}]\,,\notag\\
C &=  \sum_i [-\tfrac{1}{2}\lambda_i^{\alpha} \partial_{i \alpha} +
  \tfrac{1}{2}\tilde{\lambda}_i^{\dot{\alpha}} \partial_{i \dot{\alpha}} + \tfrac{1}{2}\eta_i^A
  \partial_{i A}]\,, \notag\\
S_{\alpha}^A &= \sum_i [-\theta_{i \alpha}^{B} \theta_i^{\beta A}
  \partial_{i \beta B} + x_{i \alpha}{}^{\dot{\beta}} \theta_i^{\beta
    A} \partial_{i \beta \dot{\beta}} + \lambda_{i \alpha}
  \theta_{i}^{\gamma A} \partial_{i \gamma} + x_{i+1\,
    \alpha}{}^{\dot{\beta}} \eta_i^A \partial_{i \dot{\beta}} -
  \theta_{i+1\, \alpha}^B \eta_i^A \partial_{i B}]\,,\notag\\
\overline{S}_{\dot{\alpha} A} &= \sum_i [x_{i \dot{\alpha}}{}^{\beta}
  \partial_{i \beta A} + \tilde{\lambda}_{i \dot{\alpha}}
  \partial_{iA}]\,,\notag\\
K_{\alpha \dot{\alpha}} &= \sum_i [x_{i \alpha}{}^{\dot{\beta}} x_{i
    \dot{\alpha}}{}^{\beta} \partial_{i \beta \dot{\beta}} + x_{i
    \dot{\alpha}}{}^{\beta} \theta_{i \alpha}^B \partial_{i \beta B} +
  x_{i \dot{\alpha}}{}^{\beta} \lambda_{i \alpha} \partial_{i \beta}
  + x_{i+1 \,\alpha}{}^{\dot{\beta}} \tilde{\lambda}_{i \dot{\alpha}}
  \partial_{i \dot{\beta}} + \tilde{\lambda}_{i \dot{\alpha}} \theta_{i+1\,
    \alpha}^B \partial_{i B}]\,.
\label{dualsc}
\end{align}
Note that if we restrict the dual generators $\bar{Q},\bar{S}$ to the on-shell superspace they
become identical to the conventional generators $\bar s, \bar q$.

\section{Some generalities on $gl(n|n)$ and its Yangian}

We will begin with the defining representation of $gl(m|n)$.
We define $E^{\AA}{}_{\BB}$ to be an
$(m|n) \times (m|n)$
matrix with a 1 in the entry in row $A$ and column $B$ and 0 everywhere else. The matrix satisfies the product
\be
E^{\AA}{}_{\BB} E^{\CC}{}_{\DD} = \delta^{\CC}_{\BB} E^{\AA}{}_{\DD},
\ee
from which follows the commutation relations of $gl(m|n)$,
\be
[E^{\AA}{}_{\BB},E^{\CC}{}_{\DD}] = \delta^{\CC}_{\BB} E^{\AA}{}_{\DD} - (-1)^{(\AA+\BB)(\CC+\DD)} \delta^{\AA}_{\DD} E^{\CC}{}_{\BB} = f^{\AA}{}_{\BB}{}^{\CC}{}_{\DD}{}_{\EE}{}^{\FF} E^{\EE}{}_{\FF},
\ee
where the structure constants $f$ are given by
\be
f^{\AA}{}_{\BB}{}^{\CC}{}_{\DD}{}_{\EE}{}^{\FF} E^{\EE}{}_{\FF} = \delta_{\BB}^{\CC} \delta_{\EE}^{\AA} \delta_{\DD}^{\FF} - (-1)^{(\AA+\BB)(\CC+\DD)} \delta_{\DD}^{\AA} \delta_{\EE}^{\CC} \delta_{\BB}^{\FF}.
\ee
If we remove the supertrace from the generators $E^{\AA}{}_{\BB}$ then we have the algebra $sl(m|n)$. In the case where $m=n$ we can also remove the trace, leading to $psl(n|n)$.

One can define a metric on $gl(m|n)$ by taking the supertrace of the product of two generators in the fundamental representation,
\be
g^{\AA}{}_{\BB}{}^{\CC}{}_{\DD} = {\rm str}[E^{\AA}{}_{\BB} E^{\CC}{}_{\DD}] = (-1)^{\AA} \delta^{\CC}_{\BB} \delta^{\AA}_{\DD}.
\ee
The inverse metric is then
\be
(g^{-1}){}_{\AA}{}^{\BB}{}_{\CC}{}^{\DD} = (-1)^{\BB} \delta^{\DD}_{\AA} \delta^{\BB}_{\CC}.
\ee
We can define `raised' structure constants as
\be
f^{\AA}{}_{\BB}{}_{\GG}{}^{\HH}{}_{\EE}{}^{\FF} = f^{\AA}{}_{\BB}{}^{\CC}{}_{\DD}{}_{\EE}{}^{\FF} (g^{-1})_{\CC}{}^{\DD}{}_{\GG}{}^{\HH} = (-1)^{\GG}(\delta^{\HH}_{\BB}\delta^{\AA}_{\EE}\delta^{\FF}_{\GG} - (-1)^{(\AA+\BB)(\AA+\EE)}\delta^{\AA}_{\GG} \delta^{\HH}_{\EE} \delta^{\FF}_{\BB}).
\ee
The representation of most interest to us is the twistor (or oscillator) representation,
\be
J^{\AA}{}_{\BB} = \cW^{\AA} \frac{\del}{\del \cW^{\BB}}.
\ee
It is simple to see that this satisfies the right commutation relations,
\be
[J^{\AA}{}_{\BB},J^{\CC}{}_{\DD}] = \delta_{\BB}^{\CC} j^{\AA}{}_{\DD} - (-1)^{(\AA+\BB)(\CC+\DD)} \delta^{\AA}_{\DD} J^{\CC}{}_{\BB}.
\ee
For multi-particle invariants we take the sum over single particle representations,
\be
J^{\AA}{}_{\BB} = \sum_i j_i^{\AA}{}_{\BB}=\sum_i \cW^{\AA}_i \frac{\del}{\del \cW_i^{\BB}}.
\ee
The Yangian generators are given by the bilocal sum,
\be
J^{(1)}{}^{\AA}{}_{\BB} = \sum_{i<j}(-1)^{\CC} [J_i^{\AA}{}_{\CC} J_j^{\CC}{}_{\BB} - J_j^{\AA}{}_{\CC} J_i^{\CC}{}_{\BB}].
\label{Yangiangenerators}
\ee
They are consistent with cyclicity (i.e. invariant up to terms which are proportional to a generator of the original superalgebra) for those algebras with vanishing Killing form \cite{Drummond:2009fd}. The simple Lie superalgebras which satisfy this condition were classified by Kac \cite{Kac:1977em} and include $psl(n|n)$. It also holds for the central extension $sl(n|n)$ but not for $gl(n|n)$. This can be seen by considering the difference of the definition (\ref{Yangiangenerators}) with that which one obtains by cyclically rotating by one step. Explicitly, the only term which is not proportional to an algebra generator is the level-one hypercharge (the supertrace of (\ref{Yangiangenerators})).

\section{Induced transformation of the minors}
\label{app-comm}
In this appendix we derive the induced transformation of the minors $\MM_p$ which we quoted in equation (\ref{Mvar}). For the convenience of the reader we repeat the result here,
\be
[\mathcal{O}^{\AA}_b, \MM_p] = \sum_{i=1}^{p-1} \cW^{\AA}_i t_{bi} \, \MM_p,
\label{Mvar-app}
\ee
where  $\mathcal{O}^{\AA}_b = \sum_{i<j}\cW_i^{\AA} \mathcal{O}_{ij} t_{bj}$.
Note that because we are calculating a commutator the $gl(n)$ operator $\mathcal{O}_{ij}$ never acts on the explicit factor of $t_{bj}$ inside $\mathcal{O}^{\AA}_b$ itself.

We should consider the cases $p \leq n-k+1$ and $p > n-k+1$ separately. In the case $p \leq n-k+1$ the minor $\MM_p$ does not `wrap' (i.e. does not involve columns from the beginning and the end of the matrix). In this case we have
\be
[\mathcal{O}^{\AA}_b, \MM_p]  = \sum_{i<j} \cW_i^{\AA} t_{bj} \mathcal{O}_{ij} \MM_p = \sum_{i=1}^{p-1} \cW_i^{\AA} \sum_{j=p}^{p+k-1} t_{bj} \MM_p^{j \ra i},
\label{OAbMp}
\ee
where we have used the form of the $gl(n)$ variation of the minors from (\ref{OijMM}).
Using the `cyclic' identity which follows from the vanishing of a totally antisymmetric object with $(k+1)$ $gl(k)$ indices,
\be
t_{a i_1} (i_2\,\, i_3 \ldots i_{k+1}) + (-1)^k t_{a i_2} (i_3 \ldots i_{k+1} i_1) + t_{a i_3} (i_4 \ldots   i_1 \, i_2) + \ldots +(-1)^k t_{a i_{k+1}} (i_1 \ldots i_k) = 0,
\ee
we find that the sum on the RHS of (\ref{OAbMp}) can be written
\be
\sum_{j=p}^{p+k-1} t_{b j} \MM_p^{j \ra i} = t_{bi} \MM_p
\ee
and so the result (\ref{Mvar-app}) holds.

In the case where $p > n-k+1$ then the minor $\MM_p$ wraps around the end of the matrix, $\MM_p = (p \ldots n\, 1 \ldots p+k-n-1)$. In this case we write instead
\be
[\mathcal{O}^{\AA}_b, \MM_p]  = \sum_{i<j} \cW_i^{\AA} t_{bj} \mathcal{O}_{ij} \MM_p =
\sum_{s=p}^n \sum_{i=1}^{s-1} \cW_i^{\AA} t_{b s} \MM_p^{s \ra i}.
\ee
Now we recall that the variation we are calculating actually sits inside the integral (\ref{invOAb}). For each term in the sum over $s$ we can therefore use the constraints $\sum_1^n t_{bl} \cW^{\AA}_l = 0$ which are imposed by the delta functions in (\ref{invOAb})\footnote{The reader may worry that one of the delta functions comes with a derivative $\partial_{\BB}$ on it. However this does not matter as the only contribution which can arise by commuting a $\cW^{\AA}$ through such a derivative is proportional to the supertrace $(-1)^{\AA}\delta^{\AA}_{\BB}$ and this can be dropped when we recall that the operator $J^{(1)}{}^{\AA}{}_{\BB}$ should have the supertrace removed.}. Only one term arises every time we do this due to the antisymmetry of the minor and we obtain
\be
[\mathcal{O}^{\AA}_b, \MM_p] = -\sum_{s=p}^n \cW_s^{\AA} t_{b s} \MM_p.
\ee
Finally we can use the delta function constraint again and find that the commutator is again of the form (\ref{Mvar-app}).

\section{Details of invariance of the gauge-fixed integral}
\label{app:derivationU}
In this appendix we want to give some more technical detail about the action of the operator 
\be
\UU_{ij} = \sum_{l=k+1}^n  t_{j l} \frac{\partial}{\partial t_{i l}}~~,~~i<j\leq k
\ee
on the minor $\MM_p$, when $1<p\leq k$. The explicit expression of the $n\times k$ gauge-fixed matrix of $t_{ai}$'s is
\beq
\left(
\begin{array}{ccccc}
{1} & {} & {} & {} & \\
{} &{} & {\ddots} & {} & \\
{} &{} & {} & {1} &\\ 
{} &{} & {} & {} & 1\\
{} &{} & {} & {} &\\
{} &{} & {} & {} &\\
{} &{} & {0} & {} &\\
{} &{} & {} & {} &
 \end{array}
\left[ \begin{array}{ccc|ccc}
{} & {} &  {} & t_{1 k+1}  & {\cdots}  &t_{1 (p+k-1)} \\{} & {} & {} \\
{} & {0}& {} & {\vdots} & {\mathrm A}  &\vdots\\
{} & {} & {} & {} & {}  &\\ {} & {} & {} \\
 \hline
{1} & {} &  {} & \vdots & {\mathrm B}  &\vdots \\
{} & {\ddots} &  {} &  &   & \\
{} & {} &  {1} & t_{k k+1} & {\cdots}  & t_{k (p+k-1)}
\end{array} \right] \ldots~~~
\right)
\eeq
where we have indicated the minor $\MM_p$ with square brackets. Its particular structure  is such that only the $\mathrm A$-part contributes to the determinant. As already mentioned in the main text,  $\UU_{ij}$ copies the $j$-th row into the $i$-th one on  the non-gauge-fixed part. Therefore, if  either $i, j \in  \mathrm A$ or $i, j \in  \mathrm B$, the result vanishes due to the antisymmetry of the minor  or to its blindness to the $\mathrm B$-part, respectively. The only non-vanishing contribution is given when $i \in  \mathrm A$ and $j \in  \mathrm B$:
\beq
\UU_{ij} \MM_p = \UU_{ij} \left[ \begin{array}{ccc|cc}
{} &{} & {} &\\
{0} &{0} & {0} & {} & {i\mathrm{-th ~row}}  \\
{} & {}  & {} & {} & {}  \\
 \hline
{1} & {} &  {} &  & {}   \\
{} & {}  &  {} &  &   \\
{0} &{0}  &  {1} &  &  j\mathrm{-th ~row}
\end{array} \right] =
\left[ \begin{array}{ccc|cc}
{} &{} & {} & {} &\\
{0} &{0} & {0} & {} & {j\mathrm{-th ~row}}  \\
{} & {} &{} & {} & {} \\
 \hline
{1} &{} & {} &  {} &   {}   \\
{} & {} &{} &  {} &      \\
{0} &{0}  &  {1} &  &  j\mathrm{-th ~row}
\end{array} \right] \equiv
\left[ \begin{array}{ccc|cc}
{} &{}  & {} &\\
{0} & {0}& {0} & {} & {j\mathrm{-th ~row}}  \\
{} & {} &{} & {} & {}  \\
 \hline
{1} &{}  &  {} &  & {}   \\
{} & {}  &  {} &  &    \\
{0} & {0} &  {1} &  &  i\mathrm{-th ~row}
\end{array} \right]
\eeq
where it is possible to write the last step as the $\mathrm B$-part does not contribute to $\MM_p$.
This result is equivalent, up to a sign, to the minor $\MM_p$ where the $j$-th column of the full matrix has been substituted by the $i$-th one:
\beq
\left[ \begin{array}{ccc|cc}
{} &{}  & {} &\\
{0} & {0}& {0} & {} & {j\mathrm{-th ~row}}  \\
{} & {} &{} & {} & {}  \\
 \hline
{1} &{}  &  {} &  & {}   \\
{} & {}  &  {} &  &    \\
{0} & {0} &  {1} &  &  i\mathrm{-th ~row}
\end{array} \right]
= -
\left[ \begin{array}{ccc|cc}
{} &{}  & {} &\\
{0}  & {0}& {1} & {} & {i\mathrm{-th~row}}  \\
{}  &{} & {} & {} & {}  \\
 \hline
{1}  & {} &  {} &  & {}   \\
{}  &{} &  {} &  &    \\
{0} & {0} &  {0} &  &  j\mathrm{-th~row}
\end{array} \right] = - \MM_p^{j\rightarrow i}
\eeq
as the gauge-fixed $t_{ai}$'s matrix has the form
\beq
\left( \begin{array}{ccc}
{} &{0} & {}\\
{0} &{1_{ii}} & 0 \\
{} & {0}  & {}  \\
 \hline
{} & {0} &     \\
{} & {0}  &     \\
{0} &{0}  &  0
\end{array}
\left[ \begin{array}{ccc|cc}
{} &{} & {0} &\\
{0} &{0} & {0} & {} & {i\mathrm{-th ~row}}  \\
{} & {}  & {0} & {} & {}  \\
 \hline
{} & {} &  {0} &  & {}   \\
{} & {}  &  {0} &  &   \\
{0} &{0}  &  {1_{jj}} &  &  j\mathrm{-th ~row}
\end{array} \right]\ldots~ \right)
\eeq
Therefore, the action of operator $\UU_{ij}$ on a minor with $1<p\leq k$ is
\beq
\UU_{ij} \MM_p = - \MM_p^{j\rightarrow i} ~.
\eeq

\end{document}